\documentclass{moriond}

\def\be{\begin{equation}}
\def\ee{\end{equation}}
\def\bea{\begin{eqnarray}}
\def\eea{\end{eqnarray}}

\usepackage{ulem}

\newcommand{\Photo}{\includegraphics[height=35mm]{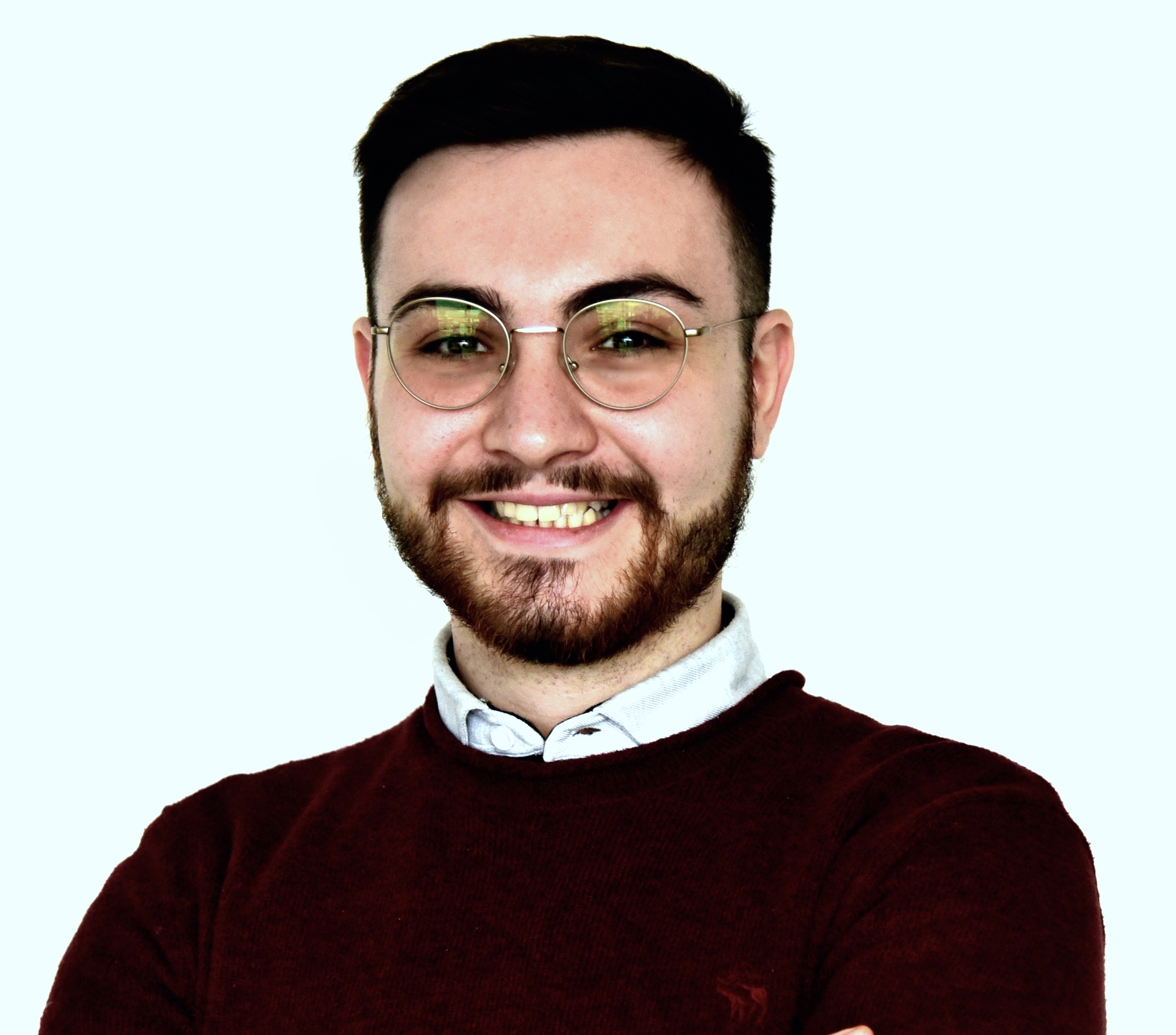}}

\begin{document}
\vspace*{4cm}
\title{Characterization of the polarized synchrotron emission from Planck and WMAP data}

\author{F. A. Martire, R. B. Barreiro and E. Mart\'{\i}nez-Gonz\'alez}

\address{Instituto de F\'isica de Cantabria, CSIC-Universidad de Cantabria,\\ 
Avda. de los Castros s/n, E-39005 Santander, Spain}

\maketitle\abstracts{The purpose of this work is to characterize the diffuse Galactic polarized synchrotron. We present EE, BB, and EB power spectra estimated cross-correlating Planck and WMAP polarization frequency maps at 23 and 30 GHz, for a set of six sky regions covering from 30\% to 94\% of the sky. The EE and BB angular power spectra show a steep decay of the spectral amplitude as a function of multipole, approximated by a power law with power indices around -2.9 for both components. The B/E ratio is about 0.22. The EB cross-component is compatible with zero at 1$\sigma$, with upper constraint on the EB/EE ratio of 1.2\% at the 2 $\sigma$ level. The recovered SED, in the frequency range 23–30 GHz, shows E and B power-law spectral indices compatible between themselves with a value of about -3.}

\section{Introduction}

An important challenge in cosmology is the precise measurement of the CMB polarization anisotropies. However, mixed with the cosmological signal, CMB observations also contain different astrophysical emissions, usually called \textit{CMB foregrounds}.
The accuracy of the CMB measurements thus depends critically on the foregrounds removal process. \\
At low frequency, roughly below 100~GHz, the dominant polarized foreground is the diffuse synchrotron emission. The synchrotron radiation is generated by relativistic cosmic ray electrons accelerating around the Galactic magnetic field, which spiral around the field lines emitting radiation. In this work, we characterize the diffuse synchrotron polarization analyzing the observations of \textit{WMAP} K-band and \textit{Planck} 30~GHz frequency channels focusing on the intermediate and low Galactic latitudes.

\section{Data and Simulations}
For our analysis, we use the \textit{Planck} \cite{Planck:2018nkj} 2018 data release (PR3), obtained from the full set of observations, focusing on the lowest channel at central frequency 28.4~GHz. Moreover, we use the lowest frequency channel of the \textit{WMAP} \cite{WMAP:2012fli} dataset, namely the \textit{K}-band centered at 23 GHz, obtained from the 9-yr data release. \\
In order to estimate the power spectra from only \textit{Planck} data, we cross-correlate the two half-ring maps. For the only-\textit{WMAP} analysis, we use as splits the co-added maps from 1 to 4 years on one side, and from 5 to 9 years on the other. By cross-correlating splits we mitigate the noise and reduce the effect of systematics. Power spectra results are also obtained from the cross-correlation of \textit{WMAP} and \textit{Planck} maps, using in this case the full-mission \textit{Planck} and the co-added nine-year \textit{WMAP} maps. By cross-correlating data from independent experiments, we can use directly the full data set rather than the splits, since the instrumental noise is uncorrelated and the effect of the systematics is also reduced. \\
The full-mission \textit{Planck} maps have significantly lower noise than the nine-year \textit{WMAP} maps, however, the synchrotron brightness in the \textit{Planck} lowest frequency, at 28~GHz, is around half that in the \textit{WMAP} K-band, at 23 GHz, what ends up in very similar foreground signal-to-noise for both experiments.

\section{Masks}
Although Galactic foregrounds studies usually focus on regions at high Galactic latitudes (since these are of greater interest for CMB analyses), the sensitivity at low frequency of \textit{WMAP} and \textit{Planck} does not allow a good characterization of the polarized signals at intermediate and high Galactic latitudes. Therefore, in order to have a higher signal-to-noise, our analysis will instead concentrate on low and intermediate latitudes, by constructing a set of customised masks with different sky fractions. \\
We define a Galactic mask in order to exclude the emission of the central part of the Galactic centre which has a very complex behaviour. 
Moreover, we masked very bright point sources, both Galactic and extragalactic, because we found that they have a significant effect at the spectra at all scales.
The combination of the Galactic and point sources masks defines a preliminary region of around 6 per cent of the sky. Once these pixels are removed, we construct a set of masks that select those areas with the largest polarization signal in the remaining 94 per cent of the sky. Thus, we mask those pixels below successively lower thresholds of $P$ in the \textit{Planck} 30~GHz polarization map. In this way, we obtain a set of 5 masks, from 30\% to 70\% sky fraction, plus the near full sky mask, 94\% sky fraction. \\
For our main results, we pick as the reference mask the one with $f_{sky}= 50\%$, which is a good compromise between the considered sky fraction and the signal-to-noise ratio.

\section{Angular Power Spectra}
In order to characterize the Galactic synchrotron polarization signal, we compute EE, BB and EB cross-spectra. We estimate pseudo-spectra with \texttt{NaMaster} \cite{Alonso:2018jzx}, including its purification method. We focus our main analyses in the multipole range 30 $\leq$ $\ell$ $\leq$ 300, binning with $\Delta\ell$ = 10 for multipoles 30 $\leq$ $\ell$ $\leq$ 200 and with $\Delta\ell$ = 20 for multipoles $\ell >$ 200. \\
We model the EE and BB synchrotron power spectra as a power law
\begin{equation}
    C_\ell^{XX} = A^{XX} \Big(\frac{\ell}{80}\Big)^{\alpha_{XX}}
\label{eq:EEBB}
\end{equation}
with $XX = EE, \ BB$. The EB cross-spectra is simply modelled as a constant
\begin{equation}
    C_\ell^{EB} = A^{EB}.
\label{eq:EB}
\end{equation}
Note that before performing the fit, the CMB contribution is subtracted from the data at the spectra level. More details about the analysis can be found in Martire et al \cite{martire}.

\subsection{WMAP-Planck cross-spectra}
\begin{figure}
\centerline{\includegraphics[scale=.45]{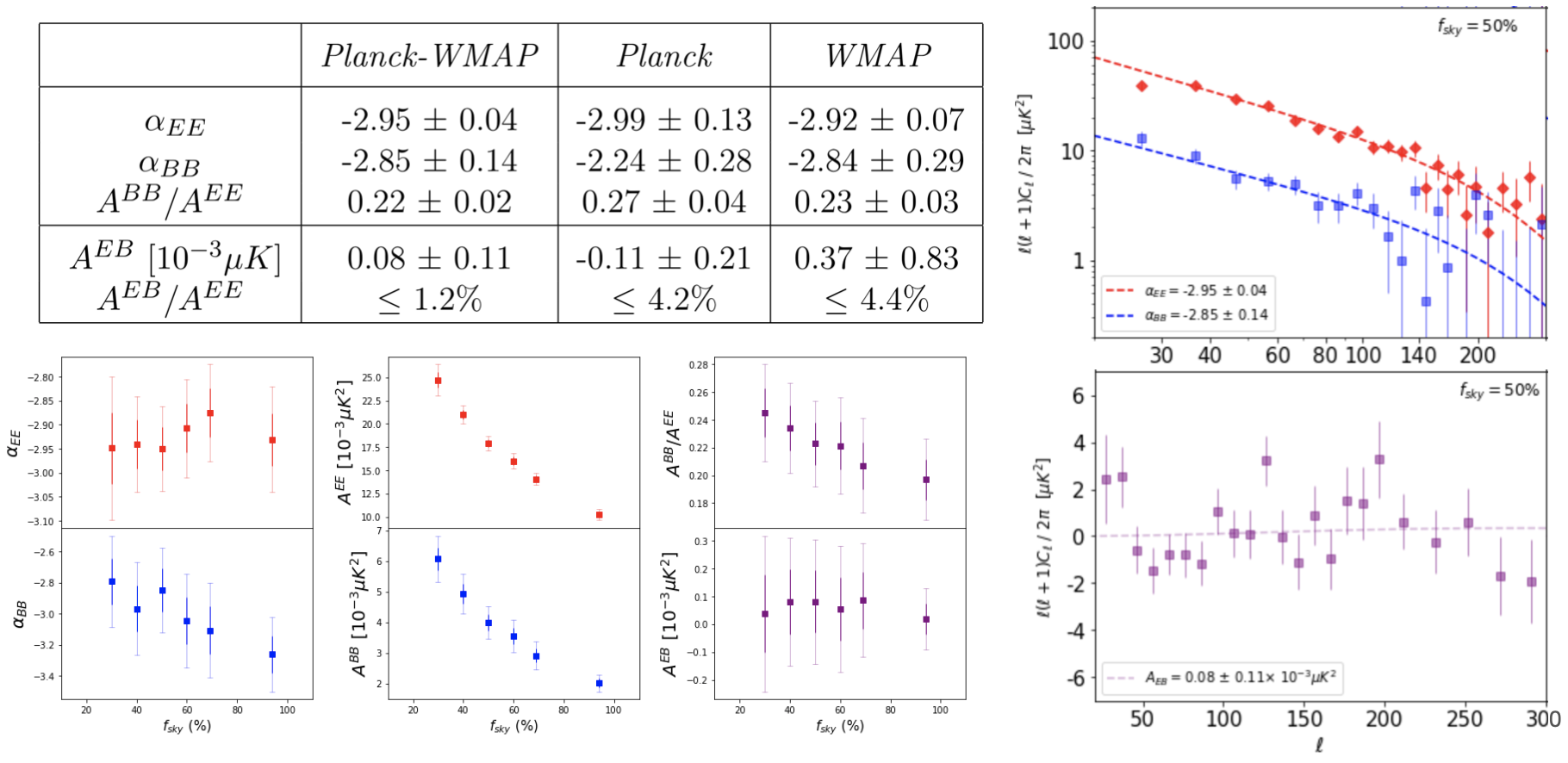}}
\caption{Left (top): Best-fit results for the reference mask ($f_{sky}=50\%$). Left (bottom): \textit{Planck}-\textit{WMAP} results for the whole mask set. Right: EE (red diamonds), BB (blue squares) and EB (purple squares) \textit{Planck}-\textit{WMAP} cross-spectra for the reference mask ($f_{sky}=50\%$)}
\label{fig:1}
\end{figure}

For our reference mask ($f_{sky}=50\%$), the EE and BB power spectra, shown in Figure \ref{fig:1} (right), of the diffuse synchrotron emission show a steep decay as a function of multipoles with consistent power spectrum indices $\alpha_{EE} = -2.95 \pm0.04$  and $\alpha_{BB} = -2.85 \pm 0.14$. Considering the results for the full set of masks shown in Figure \ref{fig:1} (left bottom), $\alpha_{EE}$ is very stable and compatible with the nearly-full sky case (94\%). Instead, the BB power law shows a slight tendency to steeper values when including high latitudes. The B-to-E ratio for the reference mask is found to be $0.22 \pm 0.02$ and it ranges from 0.20 in the 94\% mask to 0.25 for the 30\% mask. The EB cross-spectra is consistent with zero at 1$\sigma$ for the whole mask set.  We can put an upper constraint on the EB amplitude, finding it to be smaller than 1.2\% (2$\sigma$) that of the EE amplitude.\\
In order to test the robustness of our results, we fit the model to the same data set, but to a larger multipole range 10 $\leq$ $\ell$ $\leq$ 400. The conclusions agree well with the baseline results, confirming the robustness of the fits and that the model is also valid at a larger scale range.

\subsection{Planck and WMAP spectra}
The model of the synchrotron polarization spectra derived independently from \textit{Planck} and \textit{WMAP} are consistent with the power law model with null EB already obtained in the cross-analysis, as showed in Figure \ref{fig:1} (left top). However, the only-\textit{Planck} analysis suggests a less steep decay of the B-component and a slightly larger B-to-E ratio, nevertheless, the error bars overlap at 2$\sigma$ and the other analyses do not show these features. \\
As consistency test, we also performed the same analysis with the \textit{Planck} \texttt{NPIPE} \cite{Planck:2020olo} (PR4) data, cross-correlating the A/B detector splits, finding agreement with the results found with PR3.

\subsection*{Hemisphere analysis}
We repeat the same analysis independently for the Northern and Southern hemispheres, simply separating the regions from our mask set. We find that the synchrotron polarized emission in the Northern hemisphere is brighter than in the Southern hemisphere, with a factor around 1.4 larger for the amplitude of the EE spectra (slightly lower factor for BB). We also find a steeper decay of the synchrotron amplitude in the Southern hemisphere with respect to the Northern one. Nevertheless, the B-to-E ratio is quite consistent for the two hemispheres. The EB cross-term is compatible with zero at the 2$\sigma$ level for the whole mask set. \\

\section{Spectral energy distribution}
\begin{figure}
\centerline{\includegraphics[scale=.45]{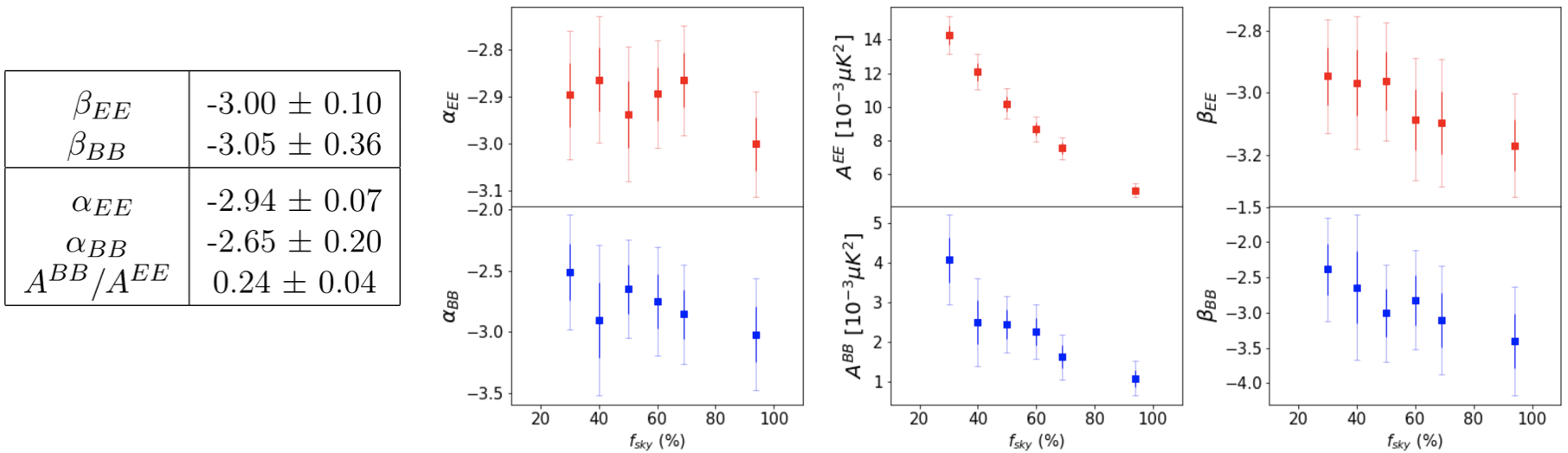}}
\caption{Left: Best-fit results for the reference mask ($f_{sky}=50\%$). Right: Results for the whole mask set.}
\label{fig:2}
\end{figure}

Using observations at different frequencies, nominally at 23 and 30 GHz, we can get information on the behaviour of the diffuse synchrotron polarization with frequency. The synchrotron spectral energy distribution (SED) is generally described by a power law $S= S_0~\nu^\beta$, where $\beta$ is the energy spectral index. If we combine it with equation \ref{eq:EEBB}, we get a system of equations that relate the energy spectral index $\beta$ and the power spectrum index $\alpha$
\begin{equation}
\left\{
\begin{array}{ll}
  (C_\ell^{XX})^{Planck}  =  (A^{XX})^{Planck} \left( \frac{\ell}{80}\right)^{\alpha_{XX}} \\
  (C_\ell^{XX})^{WMAP}  = (A^{XX})^{Planck} \left( \frac{\ell}{80}\right)^{\alpha_{XX}} \left( \frac{\nu^{WMAP}}{\nu^{Planck}} \right)^{2\beta_{XX}}
\end{array}
\right.
\label{eq:beta}
\end{equation}
with $XX = EE, \ BB$. We perform a $\chi^2$ fit to the system of equations \ref{eq:beta} for the EE and BB auto-spectra, keeping $A$, $\alpha$ and $\beta$ as free parameters. \\
For our reference mask (50\%), the spectral indices $\beta_{EE}$ and $\beta_{BB}$ are very consistent, with values of -3.00$\pm$0.10 and -3.05$\pm$0.36 respectively. Figure \ref{fig:2} shows the best-fit parameters for each of the considered masks for both the E- and B-mode components. It is interesting to point out that the power spectral indices tend to move towards steeper values when considering larger sky fractions, i.e., when including higher Galactic latitudes in the analysis.

\section*{Acknowledgments}

The authors would like to thank the Spanish Agencia Estatal de Investigaci\'on (AEI, MICIU) for the financial support provided under the projects with references PID2019-110610RB-C21, as well as from the  Unidad de Excelencia Mar{\'\i}a de Maeztu (MDM-2017-0765). We acknowledge the Legacy Archive for Microwave Background Data Analysis (LAMBDA),  supported by the NASA oﬃce of Space Science, and \textit{Planck}, an ESA science mission with instruments and contributions directly funded by ESA Member States, NASA, and Canada. This research used resources of the National Energy Research Scientific Computing Center (NERSC).

\end{document}